# 3D Protein Structure Predicted from Sequence

Debora S. Marks[1] *, Lucy J. Colwell[2] *, Robert Sheridan[3], Thomas A. Hopf[1], Andrea Pagnani[4],

Riccardo Zecchina[4], Chris Sander[3]

**\* Joint first authors**

1. Department of Systems Biology, Harvard Medical School, Boston, USA
2. MRC Laboratory of Molecular Biology, Hills Road, Cambridge, UK
3. Computational Biology Center, Memorial Sloan-Kettering Cancer Center, New York, USA
4. Human Genetics Foundation, Politecnico di Torino, Torino, Italy

**Abstract**

The evolutionary trajectory of a protein through sequence space is constrained by function and three-dimensional (3D) structure. Residues in spatial proximity tend to co-evolve, yet attempts to invert the evolutionary record to identify these constraints and use them to computationally fold proteins have so far been unsuccessful. Here, we show that co-variation of residue pairs, observed in a large protein family, provides sufficient information to determine 3D protein structure. Using a data-constrained maximum entropy model of the multiple sequence alignment, we identify pairs of statistically coupled residue positions which are expected to be close in the protein fold, termed contacts inferred from evolutionary information (EICs). To assess the amount of information about the protein fold contained in these coupled pairs, we evaluate the accuracy of predicted 3D structures for proteins of 50-260 residues, from 15 diverse protein families, including a G-protein coupled receptor. These structure predictions are *de novo*, i.e., they do not use homology modeling or sequence-similar fragments from known structures. The resulting low $C_\alpha$-RMSD error range of 2.7-5.1Å, over at least 75% of the protein, indicates the potential for predicting essentially correct 3D structures for the thousands of protein families that have no known structure, provided they include a sufficiently large number of divergent sample sequences. With the current enormous growth in sequence information based on new sequencing technology, this opens the door to a comprehensive survey of protein 3D structures, including many not currently accessible to the experimental methods of structural genomics. This advance has potential applications in many biological contexts, such as synthetic biology, identification of functional sites in proteins and interpretation of the functional impact of genetic variants.

# Introduction

**Exploiting the evolutionary record in protein families.** The evolutionary process constantly samples the space of possible sequences and, by implication, structures consistent with a functional protein in the context of a replicating organism. Homologous proteins from diverse organisms can be recognized by sequence comparison because strong selective constraints prevent amino acid substitutions in particular positions from being accepted. The beauty of this evolutionary record, reported in protein family databases such as PFAM [1], is the balance between sequence exploration and constraints: conservation of function within a protein family imposes strong boundaries on sequence variation and generally ensures similarity of 3D structure among all family members [2] (Figure 1).

In particular, to maintain energetically favorable interactions, residues in spatial proximity tend to co-evolve across a protein family [2,3]. This suggests that residue correlation can provide information about amino acid residues that are close in structure [4,5,6,7,8,9,10,11]. However, correlated residue pairs within a protein are not necessarily close in 3D space. Confounding residue correlations may reflect constraints that are not due to residue proximity but are nevertheless true biological evolutionary constraints or, they could simply reflect correlations arising from the limitations of our insight and technical noise. Evolutionary constraints on residues involved in oligomerization, protein-protein, or protein-substrate interactions or other spatially indirect or spatially distributed interactions can result in co-variation between residues not in close spatial proximity within a protein monomer. In addition, the principal technical causes of confounding residue correlations are transitivity of correlations, statistical noise due to small numbers and phylogenetic sampling bias in the set of sequences assembled in the protein family [12,13]. One does not know *a priori* the relative contributions of these possible causes of co-variation effects and is thus faced with the complicated inverse problem of using observed correlations to infer contacts between residues (Figure 1). Given alternative causes of true evolutionary co-variation, even if confounding correlations caused by technical reasons can be identified, there is no guarantee that the remaining correlated residue pairs will be dominated by residues in three dimensional proximity. The initial challenge is to solve the inverse sequence-to-structure problem by reducing the influence of confounding factors. Only then will it be possible to judge whether the evolutionary process reveals enough residue contacts, which are sufficiently spread throughout the protein sequence and structure, to predict the protein fold. The ultimate criterion of performance is the accuracy of 3D structure prediction using the inferred contacts. Previous work combined a small number of evolutionary inferred residue contacts with other, structural, sources of information to successfully predict the structure of some smaller proteins, [14,15,16,17]. However, three crucial open questions remain with respect to using evolutionary inferred residue-residue couplings for protein fold prediction. The first is whether one can develop a sufficiently robust method to identify inferred correlations that result from technical error or noise. The second is whether those remaining, truly evolutionary, inferred correlations, are dominated by residue-residue proximity, or if these can be disentangled from the whole set. The third is whether these inferred residue-residue proximities provide sufficient information to predict a protein fold, without the use of known three-dimensional structures.

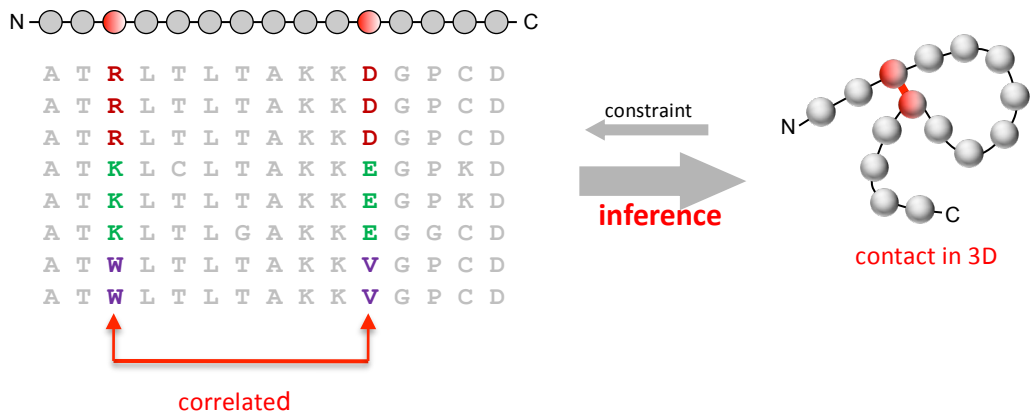

**Figure 1. Correlated mutations carry information about distance relationships in protein structure.** The sequence of the protein for which the 3D structure is to be predicted (each circle is an amino acid residue, typical sequence length is 50 - 250 residues) is part of an evolutionarily related family of sequences (amino acid residue types in standard one-letter code) that are presumed to have essentially the same fold (iso-structural family). Evolutionary variation in the sequences is constrained by a number of requirements, including the maintenance of favorable interactions in direct residue-residue contacts (red line, right). The inverse problem of protein fold prediction from sequence addressed here exploits pair correlations in the multiple sequence alignment (left) to deduce which residue pairs are likely to be close to each other in the three-dimensional structure (right). A subset of the predicted residue contact pairs is subsequently used to fold up any protein in the family into an approximate predicted 3D shape ('fold') which is then refined using standard molecular physics techniques, yielding a predicted all-atom 3D structure of the protein of interest.

**The *de novo* protein structure prediction problem in the era of genome sequencing.** Solving this inverse problem would enable novel insight into the evolutionary dynamics of sequence variation, and the role of evolutionarily constrained interactions in protein folding. Determination of protein structure, by experiment or theory, provides one essential window into protein function, evolution and design. However, our knowledge of protein structure remains incomplete and is far from saturation. In spite of significant progress in the field of structural genomics over the last decade [18], only about half of all well-characterized protein families (PFAM-A, 12,000 families), have a 3D structure for any of their members [1]. At the same time, the current upper limit on the total number of protein families (~200,000; PFAM-B) is an order of magnitude larger, and continues to grow with no clear limit in sight. Therefore, as massive genomic sequencing projects rapidly increase the number and size of protein families, in particular those without structural homologs [19], accurate *de novo* prediction of 3D structure from sequence would rapidly expand our overall knowledge of protein structures in a way difficult to achieve by experiment.

**Limited ability of current de novo 3D structure prediction methods**. Although the challenge of the computational sequence-to-structure problem remains unsolved, methods that use fragment libraries [20,21] or other strategies to search conformational space [22,23], followed by sophisticated energy optimization or molecular dynamics refinement, have been very successful at predicting the 3D structures of smaller proteins (<80 residues) [20,22,23,24] [23,25,26]. In addition, custom-designed supercomputers with orders of magnitude higher throughput have allowed insight not only into molecular dynamics of protein function, but also into the folding pathways of smaller proteins such as BPTI and WW domains [27,28]. However, none of these computational approaches have yet achieved *de novo* folding from a disordered or extended polypeptide to the native folded state for larger proteins and it is generally appreciated that the primary obstacle to 3D protein structure prediction is conformational sampling, i.e., successful search of the vast space of protein conformations for the correct fold [24,29]. Using current methods, it is computationally infeasible to adequately sample the enormous set of all 3D configurations a protein might explore in the process of folding to the native state. In this paper we explore the idea that information gleaned from statistical analysis of multiple sequence alignments can be used to solve this problem [2,5,6,30,31]. The goal is use residue-residue contacts inferred from the evolutionary record (EICs) to identify the tiny region in the space of all possible 3D configurations of a given protein that contains the correctly folded or 'native' structure.

**Extracting essential information from the evolutionary sequence record using global statistical models.** Statistical physics and computer science have developed a number of methods that address the problem of inferring a statistical model for a given set of empirically measured observables. A partial analogy can be drawn to the inverse Ising or Potts problem, in which heterogeneous local couplings between discrete state variables are derived from measurements of two-point correlation functions [32,33,34,35,36]. Similar maximum entropy methods have been applied to problems in neurobiology, e.g., for the engineering of stable and fast-folding proteins [37], for the analysis of correlated network states in neural populations [38], regulatory gene network modeling from transcript profiles [39], to extract residue-residue interactions from nucleotide sequences [40] [41]), as well as derivation of protein signaling networks from phospho-proteomics data [42]. The maximum entropy principle, which requires maximally even probabilities subject to optimal agreement between model-generated and empirical observables, turns out to be a very useful device for approaching the problem of extracting essential pair couplings from multiple sequence alignments of families of homologous proteins [43,44][11]. An alternative recently developed method, similar in intent but different in statistical approach, uses a Bayesian network framework [45] to disentangle direct from indirect statistical dependencies between residue positions and reports a dramatic improvement in the accuracy of contact prediction from multiple sequence alignments of proteins [13]. As part of the current manuscript, we evaluate whether pairs of sequence positions *i,j*, assigned high posterior probability of pairwise interaction by this interesting method (referred to here as the Bayesian Network Model (BNM)) are sufficient to predict 3D structure.

**Solving the problem of conformational complexity.** We investigate the possibility that there is enough contact information in pairwise-correlations from the evolutionary sequence record to fold a protein into a correct three-dimensional structure. Our approach builds on the ability to fit the parameters in a maximum entropy model [46] by translating and extending the resulting model to produce a set of distance constraints for effective use in distance geometry generation of 3D structures; and their refinement by energy minimization and molecular dynamics methods [47]. If this is possible, the essential data requirement for success is the availability of rich evolutionary sequence data that is sufficiently diverse to reveal co-evolution patterns in amino acid residues covering most structural elements of the protein. The goal is to use this rich evolutionary sequence information together with a global statistical model to massively reduce the huge search space of possible protein conformations.

**Testing the information content in residue co-variation about 3D structure.** We test the predictive power of this approach by generating a set of candidate structures for proteins over a range of protein sizes and different folds, including a trans-membrane protein. We quantitatively assess the extent to which predicted 3D structures have the correct spatial arrangement of α–helices and β-strands, as compared to the experimentally determined structures. We report details of blind prediction, without use of templates or fragment libraries, for 15 protein structures ranging from 48 to 258 amino acids in size and indicate how the method can be used to effectively generate rich protein structural information from sufficiently large and diverse protein family alignments. We conclude that based on our results, and the ability of high-throughput sequencing to radically augment evolutionary sequence information for different protein families, protein structure prediction from evolutionary co-variation is entirely achievable.

# Results

## Global better than local model for residue couplings

**Mutual information does not sufficiently correlate with residue proximity.** We first attempted the prediction of residue-residue proximity relationships using the straightforward local mutual information (MI) measure. $MI(i,j)$ for each residue pair $i, j$ is a difference entropy which compares the experimentally observed co-occurrence frequencies $f_{ij}(A_i,A_j)$ of amino-acid pairs $A_i, A_j$ in positions $i, j$ of the alignment to the distribution $f_i(A_i)f_j(A_j)$ that has no residue pair couplings (details in Text S1):

$$MI_{ij} = \sum_{A_i,A_j} f_{ij}(A_i,A_j) \ln\left(\frac{f_{ij}(A_i,A_j)}{f_i(A_i)f_j(A_j)}\right)$$

Eqn 1

Contact maps constructed from residue pairs assigned high *MI* values, and thus interpreted as predicted contacts, differ substantially from the correct contact maps deduced from native structures, consistent with the work of Fodor et al. [9] (Figure S1). Visual inspection of *MI*-predicted contacts as lines connecting residue pairs superimposed on the observed crystal structure confirms that the contacts predicted from *MI* are often incorrect and/or unevenly distributed (Figure 2, blue lines). Presumably this arises due to the local nature of *MI*, which is independently calculated for each residue pair *i,j*. Plausibly, the key confounding factor is the transitivity of pair correlations, where the simplest case involves residue triplets; for example, if residue B co-varies with both A and C, because B is spatially close to both A and C, then A and C may co-vary even without physical proximity (A-C is a transitive pair correlation). Any local measure of correlation, not just mutual information (MI) is limited by this transitivity effect.

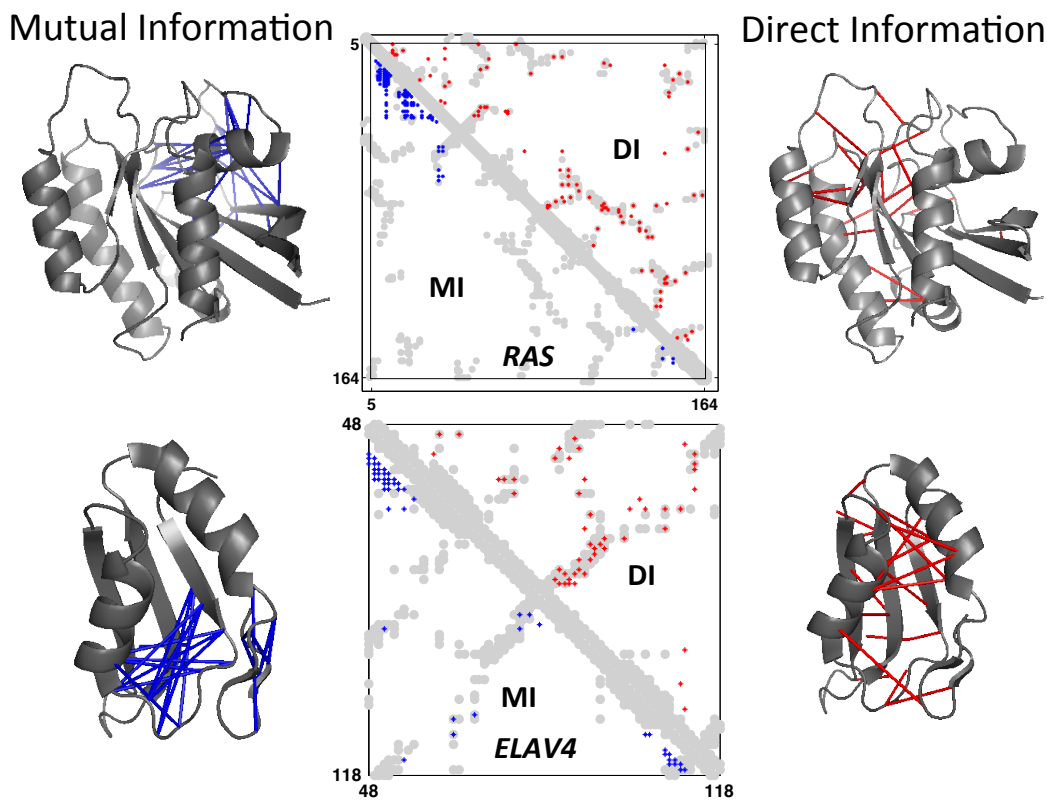

**Figure 2. Progress in contact prediction using the maximum entropy method.** Extraction of evolutionary information about residue coupling and predicted contacts from multiple sequence alignments works much better using the global statistical model (DI, right, eqn. 22 in Text S1) than the local statistical model (MI, left, eqn. 6 in Text S1). Predicted contacts for DI are better positioned in the correct structure (red lines in the observed protein structure, ribbon diagram, right), are more evenly distributed along the chain (red in contact map; upper right half) and overlap more accurately with the contacts in the observed structure (red [predicted] on grey [observed] in contact map; upper right half). Details of contact maps for all proteins in this study are available in Figures S1 and S2, Text S1.

**Effective residue couplings from a global maximum entropy model.** To disentangle such direct and indirect correlation effects, we use a global statistical model to compute a set of direct residue couplings that best explains all pair correlations observed in the multiple sequence alignment (see Methods and Text S1) [44,46]. More precisely, we seek a general model, $P(A_1...A_L)$, for the probability of a particular amino acid sequence $A_1...A_L$ of length $L$ to be a member of the iso-structural family under consideration, such that the implied probabilities $P_{ij}(A_i,A_j)$ for pair occurrences (marginals) are consistent with the data. In other words, we require $P_{ij}(A_i,A_j) \sim f_{ij}(A_i,A_j)$, where $f_{ij}(A_i,A_j)$ are the observed pair frequencies of amino acids at positions $i$ and $j$ in the known sequences in the family and the marginals $P_{ij}(A_i,A_j)$ are calculated by summing $P(A_1...A_L)$ over all amino acid types at all sequence positions other than $i$ and $j$. As specification of residue pair properties (ignoring higher order terms) leaves the amino acid sequence underdetermined, there are many probability models that would be consistent with the observed pair frequencies. One can therefore impose an additional condition, the maximum entropy condition, which requires a maximally even distribution of the probabilities - while still requiring consistency with data. Probability distributions that are solutions of this constrained optimization problem are of the form:

$$P(A_1,...,A_L) = \frac{1}{Z}\exp\left\{\sum_{1\leq i<j\leq L} e_{ij}(A_i,A_j) + \sum_{1\leq i\leq L} h_i(A_i)\right\}$$

Eqn 2

Here $A_i$ and $A_j$ are particular amino acids at sequence positions $i$ and $j$, and $Z$ is a normalization constant. The Lagrange multipliers $e_{ij}(A_i,A_j)$ and $h_i(A_i)$ constrain the agreement of the probability model with pair and single residue occurrences, respectively. This global statistical model is analogous to statistical physics expressions for the probability of the configuration of a multiple particle system, such as in the Ising or Potts models. In this analogy, a sequence position $i$ corresponds to a particle, such as a spin, and can be in one of 21 states ($A_i=1..21$); and, the Hamiltonian (the expression in curly brackets) consists of a sum of particle-particle coupling energies $e_{ij}(A_i,A_j)$ and single particle coupling energies to external fields $h_i(A_i)$.

For our protein sequence problem, the $e_{ij}(A_i,A_j)$ in Eqn. 2 are essential residue couplings that are used in the prediction of folding constraints and the $h_i(A_i)$ are single residue terms that reflect consistency with observed single residue frequencies. These parameters are thus optimal with respect to the two key conditions, (1) consistency with observed data and (2) maximum entropy of the global probability over the set of all sequences. In practice, once these parameters are determined by matrix inversion (Eqns. M4,M5), one can directly compute the effective pair probabilities $P_{ij}^{Dir}(A_i,A_j)$ (Eqn. M6), and from these the effective residue couplings ('direct information', in analogy to the term 'mutual information') $DI_{ij}$ by summing over all possible amino acid pairs $A_i,A_j$ at positions $i,j$:

$$DI_{ij} = \sum_{A_i,A_j=1}^{q} P_{ij}^{Dir}(A_i,A_j) \ln \frac{P_{ij}^{Dir}(A_i,A_j)}{f_i(A_i)f_j(A_j)}$$

Eqn 3

The crucial difference between this expression for direct information $DI_{ij}$ (Eqn. 3) and the equation for mutual information $MI_{ij}$ (Eqn. 1) is to replace pair probabilities estimated based on local frequency counts $f_{ij}(A_i,A_j)$, by the doubly constrained pair probabilities $P_{ij}^{Dir}(A_i,A_j)$, which are globally consistent over all pairs $i,j$.

**Global maximum entropy statistical model reveals residue proximity.** We now examine whether the residue coupling scores $DI_{ij}$ (Eqn 3; Eqn. 22, Text S1) from the maximum entropy model provide information about spatial proximity. Are residue pairs with higher $DI_{ij}$ scores more likely to be close to each other in 3D structure? Examination of contact maps displaying residue pairs with highly ranked $DI_{ij}$ values, overlaid onto contact maps for an observed (crystal) structure, reveals a surprisingly accurate match. The high-scoring residue pairs are often close in the observed structure, and these pairs are well distributed throughout the protein sequence and structure, in contrast to pairs with high-scoring $MI_{ij}$ values, (Figure 2, Figure S2). This remarkable level of correct contact prediction holds for all of our test cases (Table 1, Table S1) in the four main fold classes.

Others have shown that given sufficient correct (true positive) contacts combined with a lack of incorrect (false positive) contacts, predicted contacts can be implemented as residue-residue distance restraints to fold proteins from the main four fold categories with up to ∼ 200 residues to under 3Å $C_\alpha$-RMSD error from the crystal structure [48] and, in later work, up to 365 residues with accuracy under 3Å $C_\alpha$-RMSD error [48,49]. We were therefore encouraged to use our blindly predicted proximity relations as residue-residue distance restraints to fold proteins *de novo* from extended polypeptide chains.

## Protein all-atom structures predicted from evolutionary constraints

In spite of elegant analyses using subsets of real contacts [48,49], it is not *a priori* obvious to what extent accuracy of contact prediction translates to accuracy of 3D structure prediction and, in particular, how robust such prediction is to the presence of false positives. We therefore decided to assess the accuracy of contact prediction by the very stringent criterion of accuracy of predicted 3D structures.

**Generating model structures.** Starting from an extended polypeptide chain with the amino acid sequence of a protein from the family (Table S1) we used well-established distance geometry algorithms, as used for structure determination by NMR spectroscopy [50] (Box 1, Text S1). Our distance constraints were constructed using residue pairs with high DI scores pairs and secondary structure constraints predicted from sequence, Text S1, Appendix A1, Table S2. Our protocol

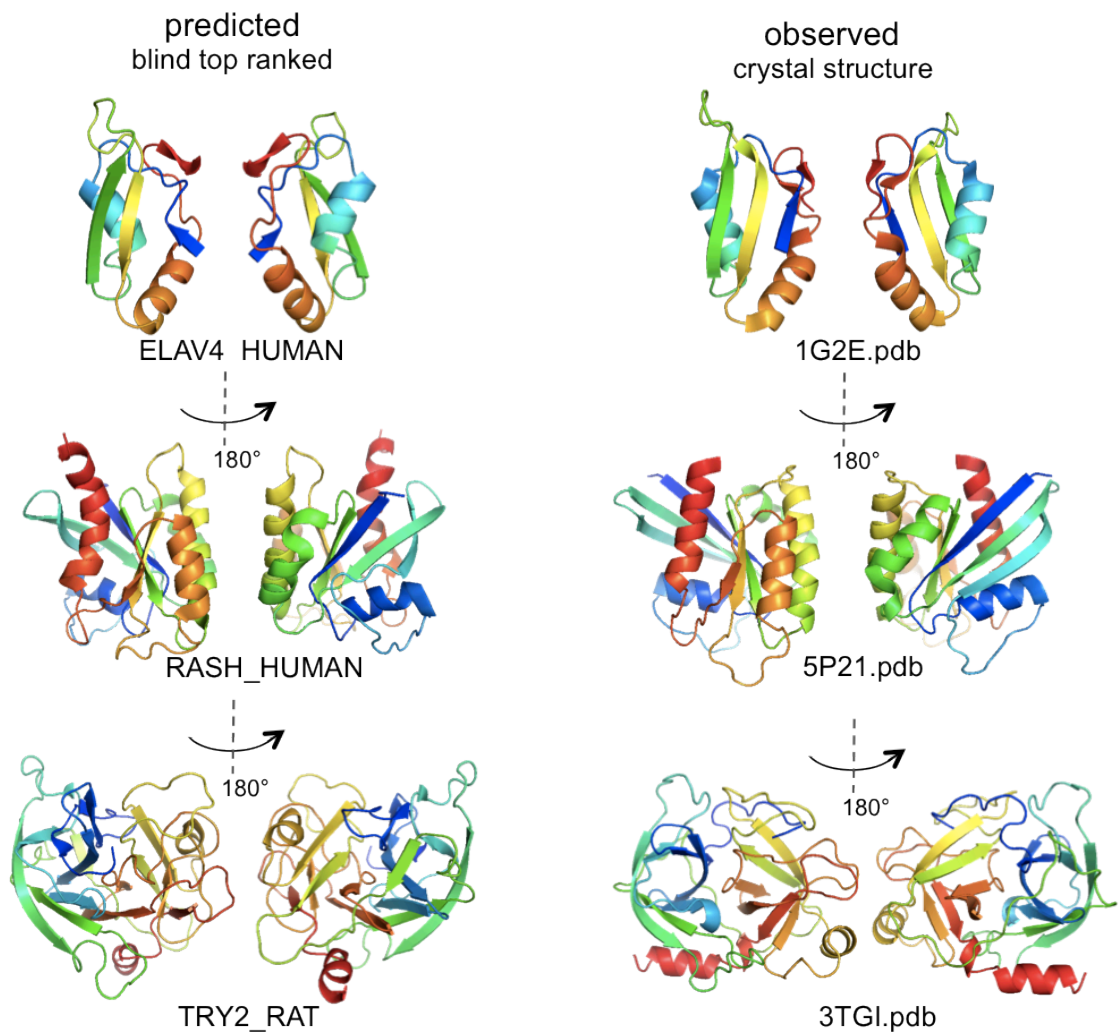

**Figure 3. Predicted 3D structures for three representative proteins.** Visual comparison of 3 of the 15 test proteins (others in Figure S3) reveals the remarkable agreement of the predicted top ranked 3D structure (left) and the experimentally observed structure (right). Center: $C_\alpha$-RMSD error and, in parentheses, number of residues used for $C_\alpha$-RMSD error calculation, e.g., 2.9Å $C_\alpha$-RMSD (67). The ribbon representation was chosen to highlight the overall topographical progression of the polypeptide chain, rather than atomic details such as hydrogen bonding (colored blue to red in rainbow colors along the chain, N-term to C-term; helical ribbons are α-helices, straight ribbons are β-strands, arrow in the direction of the chain; each structure in front and back view, related by 180 degree rotation). The predicted proteins can be viewed in full atomic detail in deposited graphics sessions for the Pymol program (Web Appendix A4) or from their coordinates (Web Appendix A).

generates initial 3D conformations to which simulated annealing is applied [47] (steps outlined in Box 1, Text S1 and Appendix A2). We reasoned that the number of distance constraints ($N_C$) needed should scale monotonically with the protein length *L*, as seen in fold reconstruction from observed contact maps [48,49]. To explore the variability of predicted structure using a given set of distance restraints, we generated 20 candidate structures for a range of $N_C$ values which started at $N_C$ =30 and incremented in steps of 10 to the nearest multiple of 10 to *L*, e.g., from *$N_C$=30* to *$N_C$=160* for
7

the Hras proteins which has 160 core residues in the PFAM alignment. Thus, in total we generate about *2\*L* candidate three-dimensional structures for each protein family as prediction candidates (Table 1, Appendix A3). Each candidate is an all-atom structure prediction for a particular reference protein of interest from the family. The model structures satisfy a maximal fraction of the predicted distance constraints and meet the conditions of good stereochemistry and consistency with non-bonded intermolecular potentials. The top structure for each protein is selected by blind ranking of these candidate structures (Figure 3, Figure S2, Appendix A3).

### 3D structure prediction for small and larger proteins is possible in diverse structural families

To evaluate the information content of residue pair correlations with respect to protein fold prediction, we apply the method to increasingly difficult cases. We start with small single-domain proteins and move on to larger, more difficult targets, eventually covering a set of well-studied protein domains of wide-ranging biological interest, from different fold classes. We report detailed results for four example families, and summary results for 11 further test families, and provide detailed 3D views of all 15 test protein families in Figure S3 and detailed 3D coordinates and Pymol session files for interactive inspection in Appendices A3 and A4, http://cbio.mskcc.org/foldingproteins

**Small: an RNA binding domain (RRM).** The blind prediction of the 71-residue RRM domain of the human Elav4 protein (Uniprot ID: Elav4_human) is a typical example of a smaller protein. The distance constraints are derived from a rich corpus of 25K example proteins in the PFAM family. The highest ranking predicted structure has a (excellent) low 2.9Å $C_\alpha$-RMSD deviation from the crystal structure over 67 out of 71 residues, a TM score of 0.57 and GDT_TS 54.6, indicating overall good structural similarity to the observed crystal structure, [51,52], (Figure 3 (top), Table 1). It has correct topography of the five β-strands and two a-helices, marred only by a missing H-bond pattern between strands 1 and 3, at least partly due to the truncation of the strand 1, a consequence of the short length of the sequence in the PFAM alignment. Strands 2 and 3 align with only 1.6 Å $C_\alpha$-RMSD deviation over the length of the predicted strands and are positioned well enough for hydrogen bonding, with some correct registration. Interestingly, the 4$^{th}$ β–strand (penultimate) missed by the secondary structure prediction method is placed in the correct region in 3D: this is one of several examples in which residue coupling information overrides incorrect local prediction. The predicted top-ranked domain of Elav4 very likely lies within the refinement basin of the native structure.

**Medium size: Ras oncogene (G-domain), an $\alpha/\beta$ domain with an GTPase active site.** The G-domain family in PFAM, with Human Ras proto-oncogene protein (Uniprot-ID: hras_human) chosen as the protein of interest, has a core MSA of 161 residues. The structure has an a/b fold with a 6-stranded β-sheet, surrounded by 5 $\alpha$-helices, one of which (a-2) is involved in the GTPase switch transition after GTP hydrolysis. The highest ranked, blindly predicted structure is 3.6Å $C_\alpha$-RMSD from the crystal structure, over 161 residues (Figure 3 (middle)) and has a high TM score of 0.7 (range 0.0 - 1.0, with 1.0 implying 100% of residues are within a set distance from the correct position,[51]) . The six β-strands and five $\alpha$-helices are placed in the correct spatial positions and are correctly threaded (Appendices A3 and A4). The 6 β-strands, which make 5 β-strand pairs are not within hydrogen boding distance for all backbone bonding, but the correct register can be easily predicted for 26/30 of the residue pairs, Text S1. The accuracy of overall topography of the highest-ranked structures is remarkable (Figure 3) and, as far as we know, currently not achievable for proteins of this size by any *de novo* structure prediction method [25].

**Larger: trypsin, an enzyme with a two-domain β-barrel structure.** The largest (non-membrane) protein family tested in the blind test is the trypsin-fold serine protease family, with rat trypsin chosen as a representative protein. Its size, at 223 amino acids, is significantly larger than proteins that can be predicted by other de novo computational methods. Trypsin consists of β-strands in two structurally isomorphous β-barrel domains. The highest-ranked predicted structure has 4.3Å $C_\alpha$-RMSD error over 186 out of 223 residues (Figure 3(bottom), Table 1, Appendices A3 and A4). The overall distribution of secondary structure elements in space is approximately correct and our method correctly predicts 5 disulfide bonded cysteine pairs, which lie within our alignment, Text S1. The topography of the first β-barrel (domain 1) is good and plausibly within refinement range of the observed structure. Five correct pairs of β-strands are identified (one absent) and 70% of hydrogen bonding paired residues are predicted with correct register, Text S1.
However, domain 2 has a number of incorrect loop progressions (see Pymol session in Appendix A3), and possibly (by inspection) is not within refinement range of the correct structure. Predicting the structure of proteins in the trypsin

**Table 1. Accuracy of predicted protein structures**

| | Target protein | | | Family containing the target protein | | Top blindly ranked predicted structure | | Best quality predicted structure | | Contact prediction quality | Known ref. structure |
|---|---|---|---|---|---|---|---|---|---|---|---|
| | Uniprot ID | Fold type | Length [# res] | PFAM ID | No. seqs | Cα-rmsd error [Å (#res)] | TM quality [0.0-1.0] | Best quality predicted structure | TM quality [0.0-1.0] | True positives for Nc=50 [0.0-1.0] | Protein Data Bank ID |
| 1 | RASH_HUMAN | α/β | 161 | Ras | 10K | 3.5 (161) | 0.7 | 2.8 (155) | 0.76 | 0.8 | 5p21 |
| 2 | CHEY_ECOLI | α/β | 114 | Response_reg | 72K | 2.98 (107) | 0.65 | 2.96 (107) | 0.67 | 0.67 | 1e6k |
| 3 | THIO_ALIAC | α/β | 103 | Thioredoxin | 13K | 3.86 (94) | 0.55 | 3.5 (97) | 0.59 | 0.68 | 1rqm |
| 4 | RNH_ECOLI | α/β | 141 | RNase_H | 11K | 4.0 (110) | 0.54 | 3.5 (114) | 0.57 | 0.68 | 1f21 |
| 5 | TRY2_RAT | β | 223 | Trypsin | 16K | 4.27 (186) | 0.6 | 4.27 (186) | 0.54 | 0.81 | 3tgi |
| 6 | CADH1_HUMAN | β | 100 | Cadherin | 12K | 3.8 (88) | 0.55 | 3.86 (96) | 0.57 | 0.86 | 2o72 |
| 7 | YES_HUMAN | β | 48 | SH3_1 | 6K | 3.6 (47) | 0.37 | 3.35 (43) | 0.41 | 0.52 | 2hda |
| 8 | O45418_CAEEL | α+β | 100 | FKBP_C | 8K | 4.1 (88) | 0.48 | 3.4 (79) | 0.53 | 0.77 | 1r9h |
| 9 | ELAV4_HUMAN | α+β | 71 | RRM_1 | 28K | 2.9 (67) | 0.57 | 3.16 (71) | 0.59 | 0.71 | 1g2e |
| 10 | A8MVQ9_HUMAN | α+β | 107 | Lectin_C | 5K | 4.8 (85) | 0.39 | 4.0 (100) | 0.53 | 0.8 | 2it6 |
| 11 | PCBP1_HUMAN | α+β | 63 | KH_1 | 9K | 4.69 (46) | 0.25 | 4.61 (61) | 0.35 | 0.47 | 1wvn |
| 12 | OPSD_BOVIN | α tm | 258 | 7tm_1 | 27K | 4.84 (171) | 0.5 | 4.29 (180) | 0.55 | 0.38 | 1hzx |
| 13 | BPT1_BOVIN | α+β | 52 | Kunitz_BPTI | 2K | 2.73 (53) | 0.49 | 2.75 (53) | 0.49 | 0.71 | 5pti |
| 14 | OMPR_ECOLI | α | 77 | Trans_reg_C | 24K | 4.7 (64) | 0.35 | 3.9 (62) | 0.45 | 0.38 | 1odd |
| 15 | SPTB2_HUMAN | α | 108 | CH(calp hom) | 4K | 4.0 (47) | 0.37 | 3.88 (88) | 0.5 | 0.5 | 1bkr |

family is particularly challenging, as the structure is known to shift s
and, as the N-terminal and C-terminal peptide cross from one doma
In spite of the limited quality of structure prediction in domain 2, it
$C_\alpha$ atoms of the highly conserved active site triad residues Ser-His-A
3Å $C_\alpha$-RMSD (and 1.3 Å *all atom*-RMSD) error of the catalytic site of

**Exploration: rhodopsin, an α-helical transmembrane protein.** Rhodo
method. This important class of membrane proteins has 7 helices a
are inferred contains many subfamilies of class A G-protein coup
rhodopsin structure (4.84Å $C_\alpha$-RMSD error from a representati
topography of the helices is accurate (TM score 0.5), with most c
which a

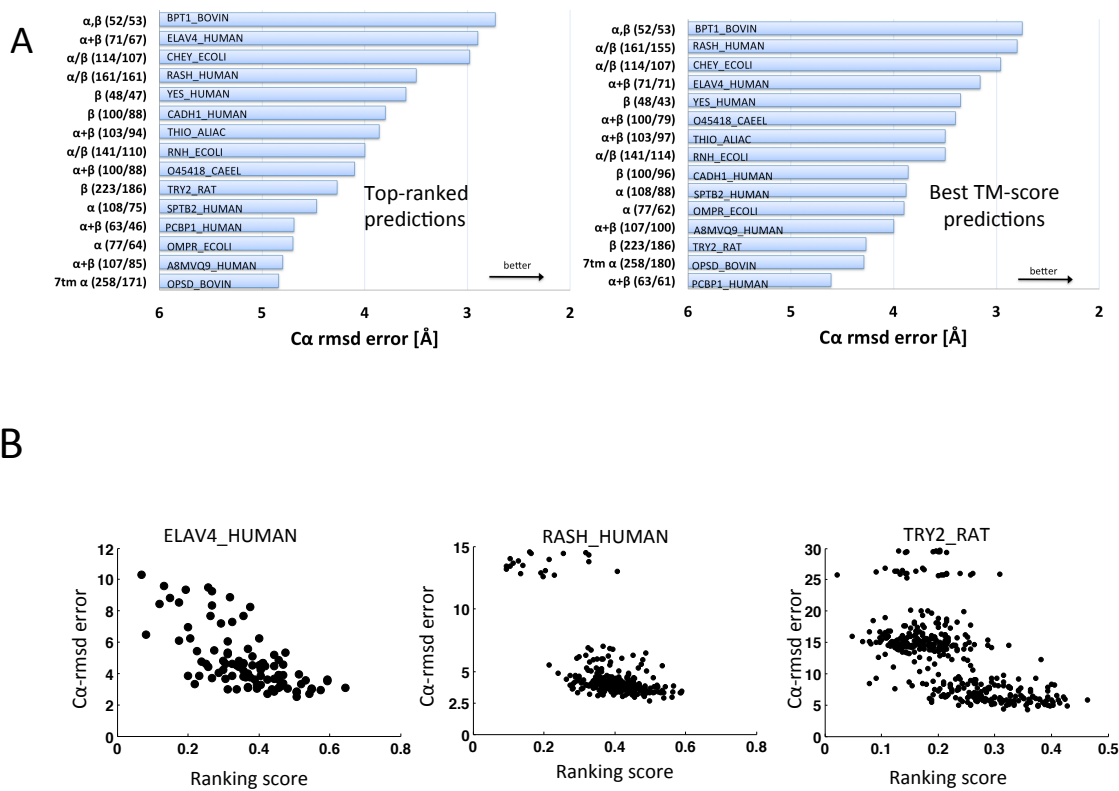

**Figure 4. Accuracy of blinded 3D structure prediction. A.** The overall performance of the *de novo* structure prediction reported here based on contacts inferred from evolutionary information (EICs), ranges from good to excellent for the 15 test proteins (on left: 3D structure type [a=α-helix-containing, b=β-strand-containing, 7tm-a = containing seven trans-membrane helices]; in parentheses: size of protein domain / number of residues used for $C_\alpha$-RMSD error calculation; on bar: Uniprot database ID). Larger bars mean better performance, i.e., lower $C_\alpha$-RMSD co-ordinate error. Left: performance for the top ranked structure for each target protein out of 2*L candidate structures in blind prediction mode; right: performance of the best structure, in hindsight, out of 20 candidate structures generated, for 20 sets of constraints ranging from 10:200, in steps of 10. This reflects what would be achievable with better ranking criteria or independent post-prediction validation of structure quality (Table 1; details of blind ranking scores in Web Appendix A5). Other well-accepted methods for error assessment, such at GDT-TS (global distance test - total score) and TM (template modeling score) are useful for comparison purposes (Table S1, Web Appendix A6).
**B**. Weighted score of each candidate structure vs $C_\alpha$-RMSD. The distribution of the 2*L candidate structures (black dots) for Elav4, Ras and Trypsin shows, in retrospect, that the ranking criteria used here are relatively useful and help in anticipating which structures are likely to be best (plots for all tested proteins available in Figure S5). In blind prediction mode, a list of predicted candidate 3D structure has to be ranked by objective and automated criteria, with a single top ranked structure or a set of top ranked structures nominated as preferred predictions. Perfect ranking would require that the highest ranked structures have the best agreement with the observed structure

The predicted structure with the highest TM score (0.55), and 4.29Å C$_\alpha$-RMSD over 180 residues, also misaligns the terminal helices but does recapitulate a network of close distances (< 4.5Å) between the side chains of Arg135 (helix III) and Glu247, Thr251 (helix VI) as well as other well-known inter-helical proximities such as Asn78 (helix II) to Trp161 (helix IV) and Ser127 (helix III) [55]. Given that the current version of the method has no information about membrane orientation for membrane proteins, this constitutes an excellent starting point for future application of this method to 3D structure prediction for membrane proteins.

**Ranking predicted structures**. To arrive at useful and objective blind predictions, the set of predicted structures for each family is ranked by objective criteria based on physical principles and a priori knowledge of general principles of protein structure. In the current implementation, we use consistency with the well-established empirical observation of right-handed chain twist in $\alpha$-helices and right-handed inter-strand twist for $\beta$-strand pairs [56] (Text S1). The virtual dihedrals of the $\alpha$-helices and the predicted $\beta$-twists in the candidate structures were combined together as a score, weighted by the relative numbers of residues in $\beta$-strands and $\alpha$-helices for each protein, see scores for all structures in Appendix A5. We found these geometric criteria effective in eliminating artifacts that appear to arise from the fact that distance constraints do not have any chiral information, such that the starting structures prior to MD refinement, while consistent with distance constraints, may have incorrect chirality, either globally or locally. We also eliminated candidate structures with knots (as with the top ranked trypsin prediction) according to the method of Mirny et al. [57].

The highest-ranked all-atom model structure is taken as the top blindly predicted structure (Table 1, Table S1). Lower ranked structures are expected to have lower accuracy of 3D structure, but this has to be tested after blind prediction by comparison with known structures. As a test of the entire procedure and the ranking criteria, we assessed our blind predictions by comparing the ranking score of the predicted structures with the experimentally observed structure, from X-ray crystallography, of the chosen reference protein (PDB), (Text S1, Figure 4A, Figure S5 and Appendix A5). For proteins such as RAS and Trypsin (Figure 4B), the objective criteria successfully ranks those predicted structures with the lowest C$_\alpha$-RMSD error to a crystal structure as highest scoring. As we remove obviously knotted proteins [57] we would miss genuinely knotted proteins [58] which are however, rarely observed.

**Assessment of prediction accuracy: 3D structures**

**Summary of blinded 3D accuracy for 15 test proteins of known structure.** We were surprised at the extent and high value of the information in the derived distance constraints about the 3D fold of examples from all major fold classes containing various proportions of $\alpha$-helices and $\beta$-sheets.
This high information content in residue couplings, derived from the maximum entropy statistical model, extends, so far, to proteins as large as G-domains, like H-ras, with 161 residues, and serine proteases, like trypsin, with 223 residues, as well as the rhodopsin family, a trans-membrane protein, with 258 aligned residues. This size has so far been out of range for state-of-the-art *de novo* prediction methods even when three-dimensional fragments are used [20,59]. In general we find that predicted $\alpha/\beta$ folds, among the 15 proteins investigated in detail, produce the most accurate overall topography (Table1, Table S1, Figure S5.). We anticipate that these results will likely extend to many protein families and that accurate structures can be generated for many of these using distance constraints derived from evolutionary information and predicted secondary structure alone, followed by energy refinement. For 12 out of the set of 15 protein families (Table 1), the top blindly ranked structures have coordinate errors from 2.7Å-4.8Å for at least 75% of the residues (using the accepted practice of omitting a moderate fraction of badly fitting residues in order to avoid exaggerated influence from outliers resulting from the square in the definition of C$_\alpha$-RMSD; comparisons carried out using the MaxCluster suite [60]). For most practical purposes, one might consider these to be within the basin of attraction within which one is highly likely to be able to identify the particular correct fold, which we estimate roughly to have a radius of about 5Å C$_\alpha$-RMSD. The partial exceptions are rhodopsin (OPSD) for which the relatively low 4.8Å C$_\alpha$-RMSD error is limited to 171 out of 258 residues (66%); and PCBP1 at 4.7Å for 46/63 residues (73%). For these proteins, the agreement is limited to a smaller, though still sizable, fraction of the protein and it is less likely that the correct overall fold would be recognized. The major exception is SPTB2 at 4.0Å for 47/108 residues (44%), which we consider not satisfactory. The TM scores customary in CASP reflect these differences and it is plausible that the top-ranked predictions for 11 out of the 15 test proteins would be considered excellent for de novo modeled structures of this size (Table S1) [25,59,61].

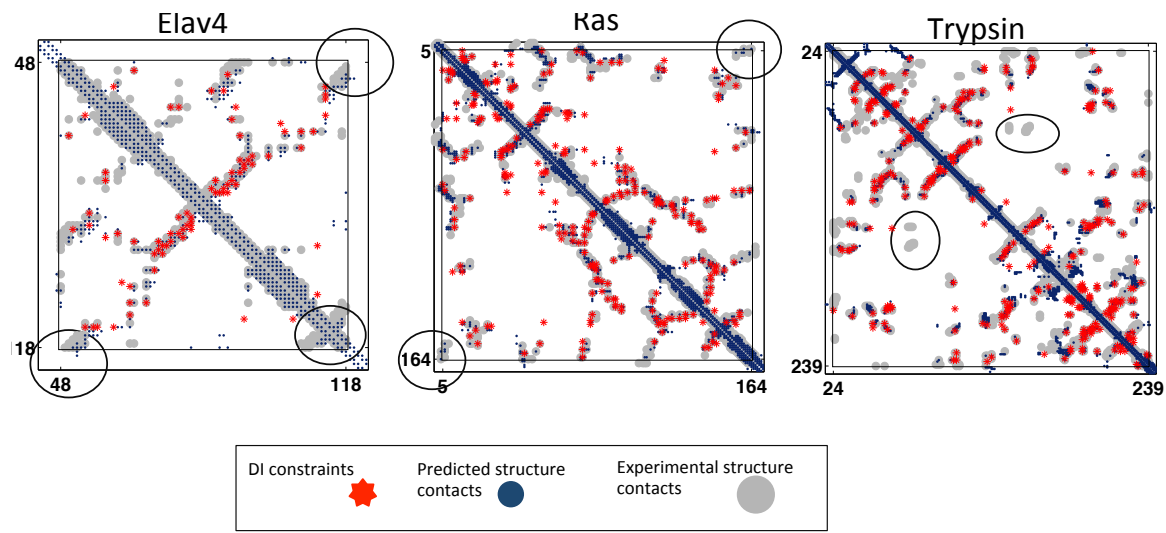

**Figure 5. Top ranked predicted structures can make correct contacts in the absence of constraints and avoid incorrect contacts in spite of false positive constraints.** The top blindly ranked structures are evaluated in terms of quality of contact prediction ($N_C$=40 for Elav4, $N_C$=130 for Ras, $N_C$=160 for Trypsin). The predicted constraints (red stars) are correct when they coincide with contacts derived from the observed structure (grey circles and otherwise incorrect (false positives, red on white). The contacts derived from the predicted structure (dark blue) are in good general agreement with those from the observed structure (grey). The cooperative nature of the folding prediction process permits favorable situations, in which contacts regions not touched by a predicted constraint (red) are still predicted correctly (black circle for RAS, dark blue on grey, no red) and false positive constraints are not strong enough to lead to incorrect contacts (left black circle Elav4, red star, no dark blue or grey). However, in unfavorable situations missing constraints may imply that contact regions are fully or partially missed (black circle, trypsin) and mostly missed (right black circle for Elav4, grey adjacent to and wider than dark blue).

Detailed examination of the close contacts of top ranked predicted structures reveals interesting violations, (Figure 5). For Ras and Trypsin false positive DI constraints (between Ser145 and Asp57 for Ras, and Ser127 and Ala37 for trypsin) are not satisfied in the top predicted structures thereby improving the accuracy. Conversely, a contact is made between the N-terminal β-strand and the C-terminal helix in RAS and C-terminal β-strand in ELAV4, despite the fact that no constraints are used in the vicinity of these contacts (grey circles, Figure 5).

**Best 3D prediction accuracy in top 400 candidate structures.** To assess the potential of the method and with a view toward future improvements of ranking criteria for sets of candidate structures, one can ask the question, from hindsight, which of, say, 400 candidate structure has the highest accuracy. This question is analogous to protein structure prediction reports that discuss the relationship (scatter plots) of, e.g., model energy against model error. Here, the best candidate structures by TM score, selected from among 400 candidate structures for each protein ($N_C$=10-200), have TM scores from 0.5 to 0.76 and typically a lower error than the blindly top ranked structure, ranging from 2.8Å to 4.6Å $C_\alpha$-RMSD for all 15 families, covering at least 80% of the residues, with the exception of OPSD where we achieve 4.3Å for 180/258 residues (70%), (Figure 4B, Table 1, Table S1). The fact that in most cases better 3D structures are found in the top 400 candidates is a non-trivial positive indication, as the conformational search space of protein folds is so large, that random methods, or moderately effective methods, would have an exceedingly low probability of achieving errors in this low range in as few as 400 structures. However, some of the structures generated here among the top 400 appear topologically incorrect, with the polypeptide chain passing through loops in a way that is, according to visual intuition, atypical of fully correct structures. Such topologically incorrectly structures would not be within a basin of attraction of conventional energy refinement, e.g., by simulated annealing. This indicates that neither low $C_\alpha$-RMSD as a measure of overall accuracy, nor the more recently developed template modeling (TM) score, nor the GDT-TS score, are fully informative indicators of structure quality. These classic structure comparison metrics need to be supplemented by more sophisticated measures, which quantify topographical differences in chain progression in 3D space, a direction for future work [62,63], together with an analysis of violations of constraints in the spirit of Miller et al. [3]. In any case, the encouragingly high accuracy of the folds we generate amongst a relatively small number of candidates imply that improved ranking criteria may lead to a better set of top-ranked, fully blinded predictions.

**Current technical limits of 3D prediction accuracy.** As an estimate of the accuracy maximally achievable by this method and its particular implementation, we performed reference calculations using artificial, fully correct, distance constraints derived from the experimentally observed structure With this ideal set of constraints, we can construct protein structure models at an error of not lower than 1.9 – 4.2Å $C_\alpha$-RMSD (Text S1, Table S3). This places a lower bound on the expected error, inherent in the distance geometry and refinement part of the method and this error will scale to some extent with the length of the protein as others have noted [48]. We expect to iterate our procedure with proteins > 350 residues. That we achieve candidate structures close to these bounds with predicted distance constraints is consistent with the notion that the inferred residue couplings contain almost all the information required to find the native protein structure, at least for the protein families examined here.

## Assessment of prediction accuracy

**Accuracy of contact prediction**. The accuracy of prediction of 3D structures crucially depends on the accuracy of contact prediction and the choice of distance constraints from a set of predicted contacts. Note that residue-residue proximity is a different requirement than residue-residue contact, as residues may be near each other in space without any of their atoms, being in inter-atomic contact (defined as inter-atomic distance near the minimum of non-bonded inter-atomic potentials ('van der Waals'), say, about 3.5Å). Here, we use the term inter-residue contact interchangeably with inter-residue proximity, i.e. minimum atom distance of less than 5 Angstroms. We assess the accuracy of contact prediction in terms of the number of true positives and false positives among predicted contacts, i.e., those that agree and those that disagree with the contacts observed in known 3D protein structures.

We find that the highest scoring pairs provide remarkably accurate information about residue-residue proximity (Figure 6A, Box 1, Figures S6 and S7). For example, the rate of true positives is above 0.8 for the first 50 pairs for HRAS and still above 0.5 for the first 200 pairs; for other proteins, it is lower but still relatively high, e.g., above 0.7 and 0.4 for the first 50 and 200 for ELAV4. These results are consistent with our parallel evaluation of contact prediction accuracy for a large number of bacterial protein domains [46] and represent a significant improvement over local methods of contact prediction from correlated mutations or co-evolution. Not surprisingly, there is a general trend for a higher rate of true positive contact prediction to results in better predicted 3D structures, The predicted structures of proteins such as Ras and CheY with a high proportion of true positive predicted contacts tend to be more accurate than those with lower rates, for example the KH domain of PCBP1 and the calponin homology domain of SPTB2. However, this relationship between the proportion of true positives and the accuracy of the best-predicted structures is not as simple as one might have expected, Figures S6, S8 and S9. For instance the thioredoxin predicted structures are on the whole more accurate than the predicted the lectin domain (A8MVQ9_HUMAN) structures despite the fact that thioredoxin has a lower true positive rate than lectin domain for its predicted contacts. Since the quality of 3D structures could depend also on the distribution of the contacts through the chain, for each protein we also calculated the distance of a experimental contact to the nearest predicted contact and this 'spread' showed a good correlation with the RMSD accuracy achieved, (Figure S10 and Text S1).

**Comparison of contact prediction accuracy between global and local models**. How well do other contact prediction methods work? The two global models, the Bayesian network model (BNM, [13,45]) and the direct information model (DI, this work and [44] have a consistently high rate of correctly predicted contacts (true positive rate) among the top $N_C$ ranked residue pairs; in comparison two local models, mutual information (MI, Eqn. 1) and SCA [64], both have a lower rate of true positives (Figure 6A and Figure S6). The relatively high accuracy of contact prediction in the BNM model encouraged us to generate predicted 3D structures based on the BNM ranked residue pairs as the basis for inferred distance constraints, following the protocol developed for the DI model. For ten test proteins, folded all-atom 3D structures for BNM agree well with the observed structure (green structures in Figure 6B and data not shown). On the whole, the $C_\alpha$-RMSD errors are somewhat higher for the structures from the BNM model than those for the DI model (red structures in Figure 6B). In particular, using the notation [protein identifier / error for BNM / error for DI], we have: [RASH/5.6Å/2.8Å], [ELAV4/3.8Å/2.6Å], [YES/4.6Å/3.6 Å] [CADH/4.7Å/3.9Å] and trypsin did not reach an accuracy lower than 12Å $C_\alpha$-RMSD with the BNM constraints (Figure 6B and data not shown). On the other hand, the BNM and the DI

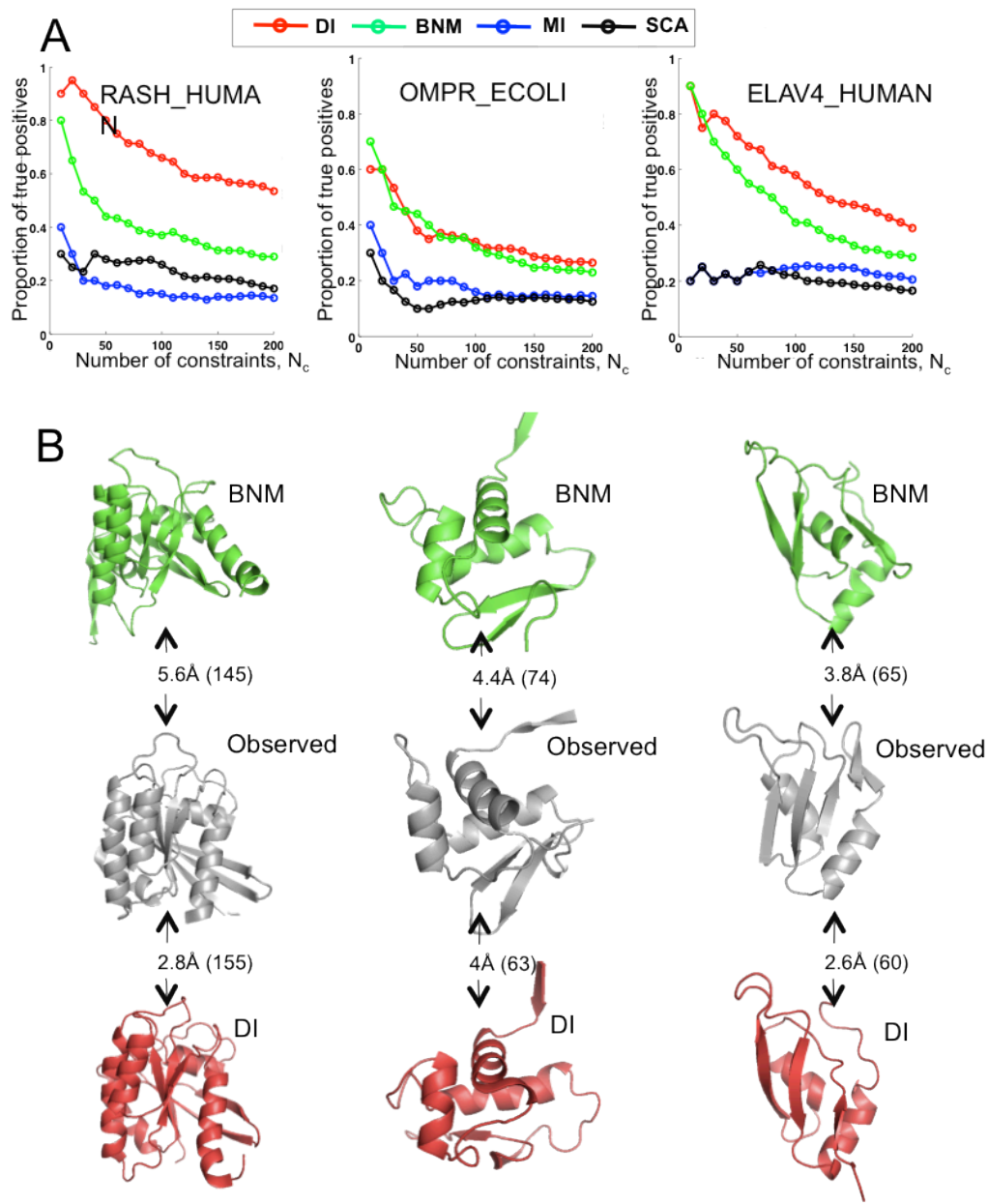

**Figure 6. Key requirement of global statistical model for correct prediction.** Evaluation of accuracy in terms of predicted contacts (A) and predicted 3D structures (B). In (A), The two global models, the Bayesian network model (BNM, green [13]) and direct information model (DI, red, this work and [46]) have a consistently high rate of correctly predicted contacts (true positive rate) among the top $N_C$ ranked residue pairs; two local models, mutual information (green, Eqn. 1 ) and SCA [64] have a consistently lower rate of true positives. Here, local refers to statistical independence of each pair i,j, while global refers to statistical consistency of all pairs. In (B), only the predicted 3D structures (green, BNM; red, EIC) for the global models have good agreement with the observed structure (grey), $C_\alpha$-RMSDs shown and residues numbers in brackets, see Web Appendix A4 for Pymol sessions containing all structures. Attempts to generate 3D structures for the two local models failed (structures not shown). Comparing (A) and (B) confirms that a higher rate of true positives for contact prediction leads to better 3D structure and that for these methods one needs at least a true positive rate of about 0.5 and on the order of about 100 predicted contacts, depending on size and other details of particular protein families. Interestingly, a false positive rate as high as about 0.3-0.5 can still be consistent with good 3D structure prediction.

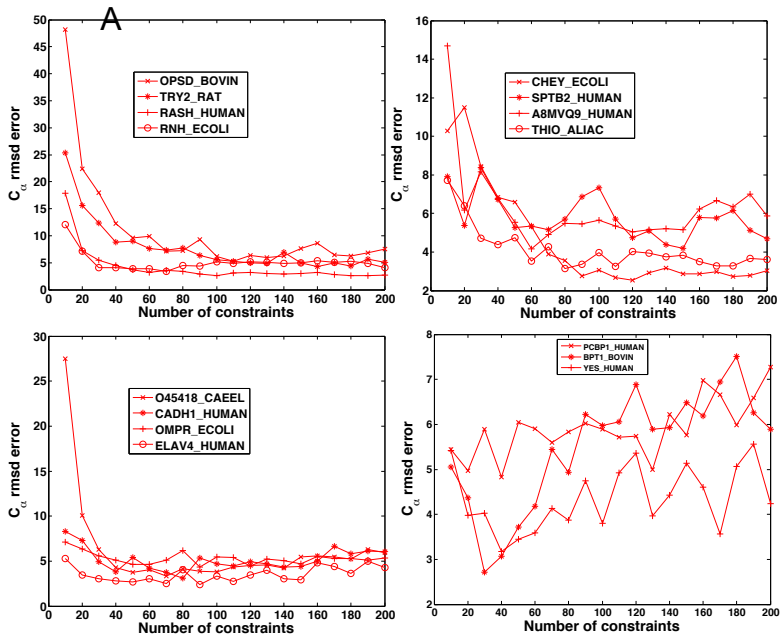

**Figure 7. Moderate number of distance constraints and varying number of sequences required for correct 3D structure prediction.**

**A. How many distance constraints are needed for fold prediction?** What fraction of false positives can be tolerated? With increasing number of predicted essential distance constraints ($N_C$, horizontal axis), 3D prediction error decreases rapidly, as assessed by $C_\alpha$-RMSD between the best of 20 (in each $N_C$ bin) predicted structures and the observed structure (here, for the 15 test proteins, using Pymol). Remarkably, as few as $\sim N_{RES}/2$ ($\sim L/2$) distance constraints $d_{ij}$ (with chain distance $|i-j|>5$) suffice for good quality predictions below 5Å $C_\alpha$-RMSD, where $N_{RES}$ is the number of amino acid residues in the protein multiple sequence alignment. We therefore routinely generated candidate protein structures for up to $N_C=N_{RES}$ distance constraints for blinded ranking (and for up to $N_C=200$ for other tests).

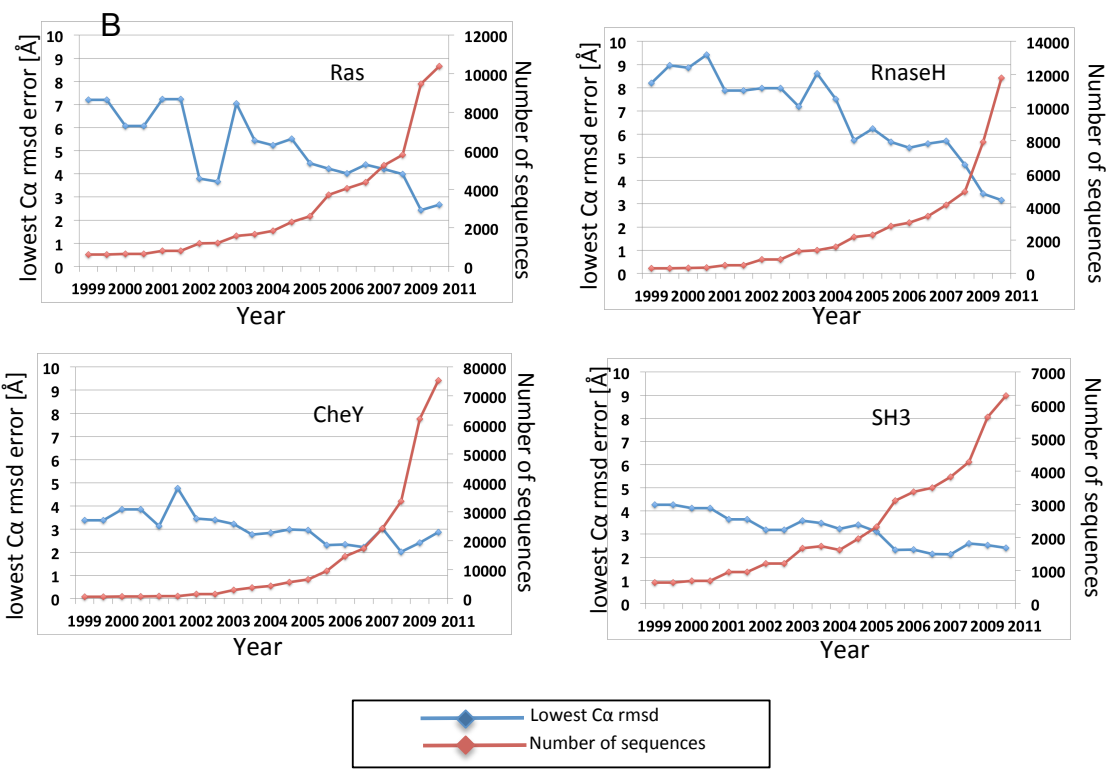

Eventually the number of false positives does degrade prediction quality, e.g., for the 58 residue protein BPTI once $N_C$ is about 80 (1.5 $N_{RES}$) the prediction quality is lost. In practice, we do not recommend using $N_C > N_{RES}$, i.e, more than about one constraint $d_{ij}$ with $|i-j|>5$, per residue.

**B. When would it have been possible to fold from sequence?**

The increase in the number of sequences available in public databases (here, from successive archival releases of the PFAM collection of protein family alignments) is one of two key elements in the ability to predict protein folds from correlated mutations. Nevertheless plotting the numbers of sequences and dates shows that it would have been possible to calculate the structures up to 10 years ago for some proteins and that amazingly few sequences are sufficient. For example, although the retrospective prediction error (vertical axis, $C_\alpha$-RMSD, using pymol) for the best 3D structure (of 400 candidates each) in four protein families (Ras, SH3 domain (YES_human) and RnaseH from Ecoli ) has decreased over time, the decrease is not strictly monotonic, as the result of non-systematic growth of the database. The point at which a predicted protein structure from a particular family reaches below 4Å $C_\alpha$-RMSD varies considerably. For example, while RnaseH required about 6000 sequence to dip below 4Å error, reached around 2008, the structure of CheY could have been predicted to 3.3Å $C_\alpha$-RMSD, with only the 600 sequences available in 1999.

predictions for OMPR were in the same accuracy range when compared to the experimental structure, as the BNM result was over 74 atoms as opposed to 63 atoms for the DI method [OMPR/4.4Å/4.0Å].

These results confirm that in general a higher rate of true positives for contact prediction leads to better 3D structure prediction; and, that for the global methods one needs at least a true positive rate of about 0.5 and on the order of about 100 predicted contacts, depending on size and other details of particular protein families. Interestingly, a false positive rate as high as about 0.3-0.5 can still be consistent with good 3D structure prediction. Clearly, the global statistical models provide a substantial increase in the accuracy of prediction of residue contacts and of 3D structures.

## Information requirements for improved prediction of 3D structures

**Requirement of sufficient sequence range coverage by the multiple sequence alignment.** Among the test set of twelve protein families, the lowest accuracy was obtained for the SPBT2 and rhodopsin proteins, (see Table 1, Table S1, Figure S3). In these cases a significant number of key residues are not included in the PFAM hidden Markov model (HMM) and thus were excluded from our analysis. If the alignment covers only part of the structure, the statistical model of the sequence is restricted to this part of the structure and does not provide information for non-covered regions. Since regions not covered by the PFAM alignments are often at the N-terminus or C-terminus of the protein and these are in contact in many protein structures, this will significantly harm the accuracy of prediction that is possible. Our analysis also shows that prediction is less likely to be accurate even within the covered region when ends of the alignment are absent. How much additional sequence information is required to build an alignment for the entire protein sequence in each case? This question is non-trivial as the diversity sampled at each sequence position by evolution varies greatly. Indeed the strength of structural evolutionary constraints may diminish towards the protein termini, analogous to the 'frayed ends' observed in many NMR-determined structures.

**Correct folding with a surprisingly small number of distance constraints**. What is the minimum number of predicted distance constraints needed to generate an approximate 3D fold? An important parameter of our folding protocol is the number of inferred distance constraints, $N_C$, used to generate candidate structures. While residues with the highest ranked pair correlations are usually close in 3D structure (Figures S6 and S7) the reliability decreases with decreasing value of $DI_{ij}$. We assessed the accuracy of the predicted protein folds for 15 evaluation families as a function of $N_C$ (Figures 7A and S11, Table S1).

Going from 10 to typically 200 distance constraints, we find that the prediction error drops sharply as EIC constraints are added, until false positives gradually start to degrade the prediction quality. We conclude that one needs about 0.5 to 0.75 predicted constraints per residue, or about 25-35% of the total number of contacts, to achieve reasonable 3D structure prediction. This number is close to those reported by other groups, who used fully correct close residue pairs to impose inexact distances as constraints [48,49,65]. For instance, Elav4 (length 71) folds to below 5Å $C_\alpha$-RMSD with only 20 constraints, whilst Trypsin (length 223) takes 130 constraints. However, the number of constraints per residue to reach below 5Å $C_\alpha$-RMSD is not constant, column 15 Table S1, and proteins such as OMPR at 0.66 constraints per residues, and Ras at 0.25 constraints per residues show that this will depend on other factors, such as type of fold and false positive rates. Whilst the accuracy of structure prediction for some proteins clearly decreases as the number of false positives, for example Cadh1, Elav4 and Yes, other proteins, such as Ras and CheY stay the same or even improve in accuracy as the false positive proportion increases, (Figure S8). This result underlines the necessity of using the constraints to attempt to fold the proteins, in order to assay the quality of predicted contacts, rather than relying on true positive rates alone.

**Increasing prediction accuracy over time, but lower than expected numbers of sequences needed**
Since we not require today's standard of high performance computing, we wondered how long ago it would have been possible to make good structural predictions. How does the accuracy of predicted folds depend on the number of sequences in the multiple sequence alignment and their evolutionary diversity? To start to explore these questions we computed the accuracy of folding using distance constraints for four representative proteins, using alignments from 20 different releases of PFAM[1] covering the last 13 years. For each multiple sequence alignment we calculated 20 structures for a range of constraints from 30-200, (Figure 7B). During this period the available sequence information has increased dramatically as the result of new sequencing technology and large-scale genome projects, so we examined the best structure attained as a function of the number of sequences. Although there is a clear overall trend for the $C_\alpha$-RMSD of predicted structures to drop monotonically as the number of sequences in the family increases (for example, RnaseH, 4Å $C_\alpha$-RMSD threshold was reached in 2009 when the number of sequences reached 5000), not all protein families behave the same way. The predicted Ras structures reached under a 4Å $C_\alpha$-RMSD in 2002 with as few as 1200 sequences, then, surprisingly, rose again as more sequences were included, to finally dip to 2.5Å $C_\alpha$-RMSD in 2009. Similarly, although the predicted structures of CheY and the SH3 domain from the Yes protein improve with the number of sequences available, predicted structures had $C_\alpha$-RMSD in errors as low as 3.3Å and 4.7Å respectively in 1999, with ~600 sequences for both.

**Align evolutionary diverged sequences**

**Calculate naïve co-occurrence matrix in pairs of sequence position (i, j) for all 400 pairs of amino acids (A,B)**

$$C_{ij}(A,B) = f_{ij}(A,B) - f_i(A)P_j(B)$$

$$C_{ij}^{-1}(A,B) = -e_{ij}(A,B)_{i \neq j}$$

**Identify maximally informative pair couplings using statistical model of entire protein to infer residue-residue co-evolution**

$$P_{ij}^{Dir}(A,B) = \frac{1}{Z}\exp\{e_{ij}(A,B) + \tilde{h}_i(A) + \tilde{h}_j(B)\}$$

$$DI_{ij} = \sum_{A,B=1}^{q} P_{ij}^{Dir}(A,B) \ln \frac{P_{ij}^{Dir}(A,B)}{f_i(A)f_j(B)}$$

high ranking transitive correlation 'indirect correlations'

re-ranked correlations 'direct information' =DI

**Analyze the highest scoring pairs to produc ranked list of residue pairs which we predict to be close in 3D space. Use these pairs as predicted close "evolutionary inferred contacts" in folding calculations**

assign (resid 143 and name CA) (resid 123 and name CA) 4 4 3
assign (resid 16 and name CA) (resid 10 and name CA) 4 4 3
assign (resid 141 and name CA) (resid 82 and name CA) 4 4 3
assign (resid 129 and name CA) (resid 87 and name CA) 4 4 3
assign (resid 92 and name CA) (resid 11 and name CA) 4 4 3
assign (resid 116 and name CA) (resid 81 and name CA) 4 4 3

predicted contacts (EICs)

**Start with extended structure use distance geometry and simulated annealing with predicted constraints, EICs, to fold the chain**

**Rank predicted structures using quality measure of backbone alpha torsion and beta sheet twist**

good scores

bad scores

**Box 1. Computational pipeline for protein folding.** The multiple sequence alignment of the protein family (MSA) is typically generated by a sequence similarity search in a large database of protein sequences to collect related sequences that are likely to have similar 3D structures. Correlations between sequence positions *i* and *j* are calculated from (co-)occurrence frequencies of amino acids in single MSA columns and column pairs, $C_{ij}(A,B) = f_{ij}(A,B)-f_i(A)f_j(B)$. By inferring a minimal statistical model of full length-sequences, which is consistent with these correlations (Text S1), direct coupling strengths $e_{ij}(A,B)$ between any pairs of residues are deduced. They help to derive distance constraints, which in turn are used to produce folded structures using these steps: distance geometry generation of approximate folds, molecular dynamics simulated annealing using standard force fields, and chirality filtering. Here, we draw any particular MSA from the PFAM collection of pre-aligned sequence families[1].

(Figure 7B). Most surprisingly, a predicted OMPR structure with an error under 5Å C$_\alpha$-RMSD would have been possibly using as few as 170 sequences (1999 PFAM release).

Hence our results highlight the overall relationship of accuracy of the predicted fold to the number of sequences available. However, this relationship is not straightforward. The distribution of sequences in the sequence space of a particular family will doubtless have an effect. In our current implementation of the algorithm, sequences with over 70% residue identity to family neighbors are down-weighted (Text S1). Therefore the effective number of sequences used for the DI coupling calculation is far less than the size of the family. Approximately only 12-40% of sequences available in the family are actually used for the calculation (Table S1). This reduction in the effective number of sequences varies substantially between families, highlighting the different distributions over sequence space covered by individual families (column 18 in Table S1). We speculate that future work will improve our understanding of *which*, as well as *how many* sequences are optimal for the contact inference from the evolutionary information.

**Discussion**

**Evolutionary constraints are determinants of 3D structure.** Protein folding algorithms tend to focus on finding the global minimum of the free energy of the polypeptide chain by physical simulations or by a guided search in conformational space using empirical molecular potentials. In this work we test the ability of a set of evolutionarily derived distance constraints between pairs of residues to guide the search towards the correct structure. As found in the study on the collective behavior of neurons, described quantitatively by models that capture the observed pairwise correlations but assume no higher-order interactions [38], our results suggest that pairwise amino-acid co-evolution statistics contain sufficient information to find the native fold. In both cases, success is contingent on the fact that indirect correlations are, at least to some extent, removed from consideration, this is achieved through the maximum entropy methodology. In the case considered here it was not necessary to explicitly consider higher order couplings, which greatly reduced the complexity of the analysis. The fact that this simplification works at all may be as much a starting point for an exploration of our understanding of the evolution of proteins as it is a route to structure prediction.

**Advantage of global statistical models.** Our calculations show that the maximum entropy approach is very effective at taking into account the interdependencies of locally calculated mutual pair information. Mutual information (MI) calculations produce many high-ranking correlations that do not reflect residue proximity. Indeed, MI tends to produce predicted contacts that are highly clustered in the contact map and have lower chain coverage, with substantial redundancy of information and a high rate of false positives from chain transitivity. In the maximum entropy calculation used to calculate the DI residue couplings, computation of the *Cij(Ai,Aj)* matrix is straightforward, given a multiple sequence alignment, however it is the matrix inversion (Eqn 18a and b, Text S1) that provides the global nature of the probability model. The application of this text-book approach from statistical physics to the problem of extracting essential pair couplings from alignments of protein sequences, with a 21-state model, leads to major progress in the problem of predicting protein-protein interactions from sequence data [11], and their use in protein folding (this work). Interestingly, an alternative approach to finding direct couplings using a Bayesian network model [BNM] also leads to improved accuracy of fold prediction using our folding protocol, compared to MI, but less so than DI couplings. A preliminary inspection showed that the overlap between the high-ranking couplings of the DI and BNM constraints is only about 40% yet the overlap contains an enhanced proportion of true positives. Understanding the theoretical connections between the two approaches may help combine the algorithms to improve the accuracy of the inferred contacts for deriving correct protein folds.

**Extracting proximity information for very conserved residues.** Completely conserved residues provide no information about pair correlations, by definition. However, the ability to predict distance constraints between highly conserved residues is a valuable feature of the DI algorithm presented here, and, in contrast to other homology-free protocols, allows direct deduction of structural information about disulfide bonds and binding sites [22]. As described above the active site residues Ser, His, and Asp in Trypsin are accurate within 1.3Å all atom RMSD of the crystal structure, (Figure S4). Even the four different loops that form the tri-nucleotide (GTP/GDP) binding site of HRAS protein, which contain well-known highly conserved amino acids boxes (GKS, DTAGQ, NKCD, SA in one-letter amino acid notation) separated by up to 100 residues in the sequence, appear in approximately the correct spatial location around the binding pocket in the highest ranking predicted structures. The striking accuracy of prediction of which loops participate in substrate sites formed by sequence-distant residues is consistent with strong evolutionary constraint in functional areas of the protein fold. The statistical model ranks co-variation signals from nearly conserved residues sufficiently highly to contribute to the correct prediction of such sites (Text S1).

**Limitations in prediction accuracy.** Clearly some protein folds are predicted more accurately than others and this may be due to a number of different factors. One clear limitation in overall accuracy is structure generation from distance constraints, using any particular protocol, as demonstrated by the folds achieved from a control set of completely correct constraints. However, use of improved molecular dynamics approaches may lower the accuracy limits of our current pipeline and we anticipate refinement of the predicted structures using iterative approaches. Among our test set, some protein folds are predicted more accurately than others due to the quality of the predicted constraints – in particular the proportion of harmful false positives. As discussed earlier, possible reasons for false positive predictions of residue couplings include: (i) statistical background noise (e.g. low statistical resolution in the empirical correlations due to an insufficient number of proteins in the family or due to global correlations from phylogenetic bias in the frequency counts), (ii) the presence of functional constraints not involving spatially close residues, such as functional constraints imposed by protein-protein or protein-ligand interactions. In this work, we reduce the noise factor by requiring at least 1000 sequences in the protein family alignment, although one may be able to reduce this limit in the future with more refined methods for taking into account the density distribution of family members in protein sequence space, as well as the organization into protein subfamilies [66]. Functional constraints, for example resulting from interactions with external partners of the protein, or alternative conformations of the same protein as in allostery, are particularly interesting and will be the subject of future analysis.

**Contribution to the current art of 3D structure prediction.** The challenge of 3D protein structure prediction depends on the extent of sequence similarity of the sequence of interest to other protein sequences whose structure is known. The difficulty of the prediction task ranges from fairly easy, if homologs of known structure are available, to very hard, when no detectable significant sequence similarity to a protein of known structure or to a known structural motif is available. Progress in this field has been expertly assessed by the pioneering community effort, the Critical Assessment of Techniques for Protein Structure Prediction (CASP), founded by Krzysztof Fidelis, John Moult and their colleagues in 1994 [67,68,69], *www.predictioncenter.org*). A series of ingenious methods have led to significant progress as reported in CASP since then, including threading, molecular dynamics, fragment-based assembly, contact prediction, machine learning, as well as methods combining several techniques [14,61,70,71] . Internet servers have also facilitated the use of these new methods and allowed ongoing critical assessment of prediction accuracy [72,73]. On this background, the goal of this work is to assess the contribution of one primary source of information, evolutionarily inferred residue couplings, to 3D structure, rather than optimizing prediction accuracy in the field of all other methods, as is done in CASP. We anticipate that in future objective assessment exercises others may want to adopt a derivative or variant of the method presented here for use in combination methods, e.g., improved contact energy in the I-Tasser simulation method [74] [17] or addition of EIC distance restraints into the Robetta server. Here, the significant information content in inferred contacts is apparent both in the assessment of prediction accuracy both for contacts (2D) as well as for all-atom structures (3D).

**Contribution to solving biological problems.** We anticipate that our method, alone or in combination with other techniques, may soon allow 3D structures with correct overall fold to be predicted for biologically interesting members of protein families of unknown structure, with potential applications in diverse areas of molecular biology. These include (1) more efficient experimental solution of protein structures by X-ray crystallography and NMR spectroscopy, e.g., by eliminating the need for heavy atom derivatives, by guiding the interpretation of electron density maps or by reducing the required number of experimental distance restraints, as elegantly demonstrated by the Baker and Montelione groups [75]. Additional interesting potential applications include (2) a survey of the arrangements of trans-membrane segments in membrane proteins; (3) discovery of remote evolutionary homologies by comparison of 3D structures beyond the power of sequence profiles [76](4) prediction of the assembly of domain structures and protein complexes [77] (5) plausible structures for alternative splice forms of proteins; (6) functional alternative conformers in cases where our approach generates several distinct sets of solutions consistent with the entire set of derived constraints; and (7) generation of hypotheses of protein folding pathways if the DI predictions involve residue pairs strategically used along a set of folding trajectories. We also anticipate that structural genomics consortia would benefit greatly from reasonably accurate predictive methods for larger proteins, for example, to (8) prioritize protein targets and define domains of interest for both crystallography and NMR pipelines.

**The need to accelerate structure determination.** Large investments continue in structural genomics, the global effort to solve at least one structure for each distinct protein family and to derive biological insight from these structures. While tremendous strides have been made in the last decade and experimental structure determination has been greatly accelerated, much less than 50% of the overall goal has been achieved to date. At the same time, the number of known protein families has increased as the result of massively parallel sequencing. Among the 12,000 well-organized protein domain families (PFAM-A collection of multiple sequence alignments), fewer than 6000 domain families have one member with a known 3D structure (from which plausible models can be built for all family members using the technique of model building by homology to structural templates). Beyond these, there are currently about 200,000 additional protein families with sequences that do not map to domains of known structure. The ability to calculate reasonably accurate structures for many of these families *de novo* from sequence information would enormously accelerate completion of the goal of structural genomics to cover the entire naturally occurring protein universe with known 3D structures. The speed advantage of the method under investigation here compared to experimental structure determination, derives from the increase of sequencing capacity by several orders of magnitude in the last decade. As we are about to reach a truly explosive phase of massively parallel sequencing, we anticipate increased coverage of sequence space for protein families by several orders of magnitude, well above the level of 1000 - 10000 non-redundant sequences for protein family and with rich evolutionary information about protein structure directly from sequence. We speculate that the utility of methods such as the one here has therefore not saturated, that predictions will become more accurate, and that applications will become broadly applicable to biological problems that can benefit from knowledge of protein structures.

**Protein folding in practice**. Our *de novo* folding protocol for a medium-size protein using evolutionarily derived constraints does not require high-performance computing and can be done in well under an hour on a standard laptop computer. One starts with a multiple sequence alignment, uses the maximum entropy model to predict a set of residue couplings from the protein family alignment, adds predicted secondary structures, derives a set of distance constraints, generates initial structures using distance geometry, refines these using molecular dynamics with simulated annealing and ranks predicted structures according to a set of empirical criteria. This first detailed report for 15 proteins in different fold classes suggests that one can predict reasonably accurate protein structures "on the fly" and that one will be able to pre-compute and make publically available arguably useful predicted structures for thousands of protein families in diverse fold classes in the near future.

**Materials and Methods**

The main steps (Box 1) in the blind prediction (1) and subsequent evaluation (2) of accuracy are: (1) selection of protein family alignments, computation of effective DCA coupling strengths in the maximum entropy model, secondary structure prediction, definition of EIC distance constraints inferred from evolutionary information and the number of constraints used and their relative weight, preparation of input to distance geometry and simulated annealing protocols, computation of a relatively small number of candidate structures, and application of empirical filters to rank predicted structures; and, (2) evaluation of prediction accuracy by computation of structural error of predicted contacts and predicted 3D structures relative to the reference PDB structure. See Text S1 for additional method details.

**(1) Computation of DCA residue pair coupling parameters in the maximum entropy model.** PFAM protein family sequence alignments were selected with a known crystal structure for at least one family member and with more than 1000 sequences in the family. Sequences in the family alignments were weighted to reduce potential spurious correlations due to sampling bias from redundant sequence information in dense regions of sequence space. A maximum entropy model was applied to identify a maximally informative subset of correlated pairs of columns across the family alignment. The statistical model describes the expected behavior of all residues up to pair terms as a joint probability distribution.

To compute the effective pair couplings and single residue terms in the maximum entropy model two conditions must be satisfied. The first condition is maximal agreement between the expectation values of pair frequencies (marginals) from the probability model with the actually observed frequencies:

$$P_{ij}(A_i, A_j) \equiv \sum_{\{A_k=1...q\}k\neq i,j} P(A_1 ... A_L) = f_{ij}(A_i, A_j) \qquad \text{Eqn M1}$$

where $A_i$ and $A_j$ are particular amino acids sequence positions $i$ and $j$. The second condition is maximum entropy of the global probability distribution, which ensures a maximally evenly distributed probability model and can be satisfied without violating the first condition:

$$S = - \sum_{\{A_i | i=1,...,L\}} P(A_1,...,A_L) \ln P(A_1,...,A_L) \qquad \text{Eqn M2}$$

The solution of the constrained optimization problem defined by these conditions, using the formalism of Lagrange multipliers, is of the form:

$$P(A_1,...,A_L) = \frac{1}{Z} \exp\left\{ \sum_{1\leq i<j\leq L} e_{ij}(A_i, A_j) + \sum_{1\leq i\leq L} h_i(A_i) \right\} \qquad \text{Eqn M3}$$

This global statistical model is formally similar to the statistical physics expression for the probability of the configuration of a multiple particle system, which is approximated in terms of a Hamiltonian that is a sum of pair interaction energies and single particle couplings to an external field. In this analogy, a sequence position $i$ corresponds to a particle and can be in one of 21 states, and a pair of sequence positions $i,j$ corresponds to a pair of interacting particles. The global probability for a particular member sequence in the iso-structural protein family under consideration is thus expressed in terms of residue couplings $e_{ij}(A_i,A_j)$ and single residue terms $h_i(A_i)$, where $Z$ is a normalization constant.

Computationally, determination of the large number of parameters $e_{ij}(A_i, A_j)$ and $h_i(A_i)$ that satisfy the given conditions is a complex task, which can be elegantly solved in a mean field approximation (Supplement and [46]) or, alternatively, in a Gaussian approximation [78]. In either approximation the effective residue coupling are the result of a straightforward matrix inversion

$$e_{ij}(A_i, A_j) = -(C^{-1})_{ij}(A_i, A_j) \qquad \text{Eqn M4}$$

of the pair excess matrix restricted to (q-1) states ($1 \leq A_i, B_j \leq q-1$),

$$C_{ij}(A_i, A_j) = f_{ij}(A_i, A_j) - f_i(A_i) f_j(A_j) \qquad \text{Eqn M5}$$

which contains the residue counts $f_{ij}(A_i,A_j)$ for pairs and $f_i(A_i)$ for singlets in the multiple sequence alignment The parameters $h_i(A_i)$ are computed from single residue compatibility conditions. Given the formulation of the probability model (Eqn. 1), the effective pair probabilities (with $\tilde{h}(A_i)$ as defined in the Supplement) are

$$P_{ij}^{Dir}(A_i, A_j) = \frac{1}{Z} \exp\left\{ e_{ij}(A_i, A_j) + \tilde{h}_i(A_i) + \tilde{h}_j(A_j) \right\} \qquad \text{Eqn M6}$$

These pair probabilities refer to the full specification of particular residues $A_i, A_j$ at positions $i$ and $j$. For the quantification of effective correlation between two sequence positions $i$ and $j$, one has to sum over all particular residue pairs $A_i, A_j$ to arrive at a single number that assesses the extent of co-evolution for a pair of positions. In analogy to mutual information,

$$MI_{ij} = \sum_{A_i, A_j} f_{ij}(A_i, A_j) \ln\left(\frac{f_{ij}(A_i, A_j)}{f_i(A_i) f_j(A_j)}\right)$$

Eqn M7

such that the direct coupling terms between columns $i$ and $j$ are given by

$$DI_{ij} = \sum_{A_i, A_j = 1}^{q} P_{ij}^{Dir}(A_i, A_j) \ln \frac{P_{ij}^{Dir}(A_i, A_j)}{f_i(A_i) f_j(A_j)}$$

Eqn M8

As there are $L^2$ values $DI_{ij}$, and one expects residue contacts of the order of magnitude of $L$, only a relatively small number top-ranked $DI_{ij}$ values (ordered in decreasing order of numerical value) are useful predictors of residue contacts in the folded protein. Given the analogy to statistical physics, the residue couplings $e_{ij}(A_i, A_j)$, on which the $DI_{ij}$ are based, can be thought of as pair interaction energies. The hypothesis, that only a fairly small subset of these pair terms are needed to determine the protein fold, is consistent with the very interesting physical notion that only subset of residue-residue interactions essentially determine the protein folding pathway. In practice, the validity of the probability formalism does not depend on the validity of this physical interpretation. We therefore proceed to use the ranked set of $DI_{ij}$ values as raw valuable material for the derivation of distance restraints for 3D structure prediction. The most computationally intensive step being inversion of a large matrix of pair terms, the $C_{ij}(A,B)$ matrix (over sequence positions $i=1,L$ and $j=1,L$; and amino acid residue types $A=1,20$ and $B=1,20$, of dimension $L^2 * 20^2$, with $L$ the length of the sequence of order 50-250 residues in the current application.

**(2) Selection of EIC distance constraints for use in the generation of all-atom structures.** The top-ranked set of $DI_{ij}$ (*direct information* terms, analogues of *mutual information* terms) are then translated to inferred contacts using consistency with predicted secondary structure, removal of predicted pairs close in sequence, and a conservation filter. The first $N_C$ residue pairs ranked according to their coupling scores are translated to distance constraints, i.e., bounds on the distances between $C_\alpha$ and $C_\beta$ residue and side chain centers between paired residues; and, as weighted distance restraints for structure refinement by simulated annealing using molecular dynamics, resulting in candidate all-atom protein domain structures.

**(3) Blinded structure prediction.** The protein polymers are folded from a fully extended amino acid sequence of the protein of interest using standard distance geometry techniques and simulated annealing with standard bonded and non-bonded intra-molecular potentials (in vacuum) using the CNS molecular dynamics software suite, with a simulated annealing MD protocol similar to those used in structure determination from NMR [47]. The elimination of mirror topologies and ranking of candidate structures is achieved by computing virtual dihedral angles using four appropriate $C_\alpha$ atoms, reflecting standard α-helical and β-strand pair handedness, and then adding the scores normalized to the predicted secondary structure content (Text S1 and Figure S5) Candidates are also filtered to remove knotted structures as defined by computation of an Alexander polynomial by the KNOT server [57].

**(4) Evaluation of prediction accuracy**. Accuracy of prediction of *residue-residue contacts* is quantified in 4 ways: (i) comparison of the EIC rank versus the minimum inter-residue distance in the crystal structure (Figure S7); (ii) comparison of the true positive rate of contact prediction versus the number of constraints (Figure S6); (iii) quantification of the severity of the false positives in a set of predicted constraints by measuring the mean of the distance in chain space to the nearest contact in the experimental structure (Figure S9); and (iv) quantification of the distribution (spread) of the contacts along the chain and over the structure of the protein, by measuring the mean of the distance from every experimental (crystal structure) contact to the nearest predicted contact (Figure S10).

Accuracy of prediction of *3D structure* is quantified in 3 ways: (i) using the TM score [51]; (ii) using GDT-TS [52]; and (iii) using the Pymol [79] 'align' routine, which reports the $C_\alpha$-RMSD for a moderately trimmed set of residues after iteratively removing the worst residue pairs from consideration as it finds an optimal superimposition of the residues in the predicted and the reference PDB structure.

**(5) Comparison to other contact prediction methods.** We calculated the four measures of contact prediction accuracy as in (4) above, for Mutual Information (MI), the Bayesian Network Model (BNM) [13,45] and the Statistical Coupling Analysis (SCA) [64] [80]. We tested all three methods for their ability to generate protein folds for a number of families, using exactly the same pipeline as for the DI constraints of this work. Folding with constraints derived from MI or SCA did not achieve reasonable accuracy with any of the tested families (data not shown). However, constraints derived from BNM were successful in generating de novo predicted structures at less than < 5Å $C_\alpha$-RMSD for 6 of the 10 tested proteins.

Additional method details are in Text S1.

## Abbreviations

MI (mutual information); DI (direct information); DCA (direct coupling analysis); EIC (contacts inferred from evolutionary information); BNM [ Bayesian Network Model}; SCA {Statistical Coupling Analysis]; 3D (three-dimensional); NMR (nuclear magnetic resonance); MD (molecular dynamics); CNS (Crystallography & NMR System, software for macromolecular structure determination); PDB (Protein Data Bank, repository of three-dimensional protein structures); PFAM (protein family database); UNIPROT (protein sequence database); MSA (multiple sequence alignment); CASP (Critical Assessment of Techniques for Protein Structure Prediction); $C_\alpha$-RMSD (root mean square distance between equivalent $C_a$ atoms, i.e., residue centers, after optimal superimposition of two structures); GDT-TS (global distance test - total score); TM (template modeling score).

## Acknowledgements

We thank Rebecca Ward, Roy Kishony, Nicholas Stroustrup, Michael Desai, Michael Brenner, Andrew Murray, Burkhard Rost, Reinhard Schneider, Alfonso Valencia, Linus Schumacher, David Shaw, Christine Orengo, Willie Taylor, Cyrus Chothia, Sarah Teichmann and John Moult for illuminating discussions and/or comments on the manuscript; and Martin Weigt for his role in the development of the maximum entropy direct coupling analysis method.

## Publication Process

The first version of the manuscript was completed on March 30, 2011.
Submitted to PLoS Biology on June 8, 2011.
Submitted to arxiv.org on October 23, 2011.

## Financial Disclosure

The Sander group has support from the DFCI-MSKCC Physical Sciences Oncology Center (NIH U54-CA143798). LJC is supported by an EPSRC fellowship (EP/H028064/1). No other financial support was received for the research. The funders had no role in study design, data collection and analysis, decision to publish, or preparation of the manuscript.